\documentstyle[aps,prl]{revtex}
\newcommand{\BE}{\begin{equation}}
\newcommand{\EE}{\end{equation}}
\newcommand{\BA}{\begin{eqnarray}}
\newcommand{\EA}{\end{eqnarray}}

\begin{document}
\draft

\twocolumn[\hsize\textwidth\columnwidth\hsize\csname@twocolumnfalse\endcsname
\title{Comment on ``Air bubbles in water: A strongly multiple scattering medium
for acoustic waves" [Phys. Rev. Lett. {\bf 84}, 6050, 2000]}
\author{Z. Ye$^1$, Y.-Y. Chen$^1$, A. Alvarez$^2$}
\address{$^1$Department of Physics, National Central University, Chungli, Taiwan 320\\
$^2$NATO Saclant Undersea Research Centre, Viale San Bartolomeo
400 19138 La Spezia Italy.}
\date{\today}
\maketitle

]

In this communication, we would like to comment on some results in
a recent Letter by Kafesaki {\it et al.}\cite{Kaf}.

In \cite{Kaf}, the authors stated ``Up to now the extensive
studies of bubbly water (or other bubbly liquids) have been
analyzed in the framework of single scattering". In fact, acoustic
scattering by bubbles has also been studied in terms of multiple
scattering both extensively and intensively in the literature.
This includes both theoretical (e.~g.
\cite{Morse,Commander,EPL,Condat,Sangani,Ye1,Feuillade,Ye2,Henyey,Ye3,Temkin})
and experimental works (e.~g. \cite{Silberman,Nicholas}). The work
of Foldy\cite{Foldy}, Lax\cite{Lax}, Waterman {\it et
al.}\cite{Waterman}, and Twersky\cite{Twersky} serves as a solid
foundation to the multiple scattering of acoustic waves by many
scatterers in general. The textbook by Ishimaru\cite{Ishimaru}
presents an excellent account of various theories about the
multiple scattering. The authors further suggested ``Thus,
acoustic waves in bubbly liquids (especially water), in addition
to being very important {\it per se}, provide an almost ideal case
for examining in detail the localization question". We note that
the research on multiple-scattering induced acoustic localization
in random bubbly water is rich (e.~g.
\cite{EPL,Condat,Ye3,Weston,Tolstoy,Ye4,AAD1,Ye5}). It has been
shown\cite{Ye3,AAD1,Ye5} that acoustic waves can be localized
within a range of frequencies slightly above the natural frequency
of an air-bubble in water, and the localization behavior is
insensitive to the configuration of bubble clouds\cite{Ye4}. When
waves are localized in bubbly water, a global coherent behavior
appears. This is a distinct feature differentiating the
localization effect from residual absorption\cite{Ye5,Emile,Ye8}.
The localization disappears when the air-bubble volume fraction is
lower than about $10^{-5}$, in agreement with an earlier
prediction\cite{EPL}.

To the end of their Letter, Kafesaki\cite{Kaf} stated ``The most
significant and novel results we obtained are shown in the panel
of Fig.~3(b)...". The results shown in their Fig.~3(b) are a
natural extension of the results of \cite{Ye3}. The formulation in
\cite{Ye3} stems from the work of Foldy\cite{Foldy} and
Twersky\cite{Twersky} and has been documented in detail in
\cite{Ye6,AAD2}. The approach has also been compared favorably
with others in certain situations\cite{Ye6,AAD2,Ye7,Gau}, and the
results agree with the physical intuition (e.~g.
\cite{Strasburg}). Though not exact, the approach is applicable
for most frequencies considered in \cite{Kaf}. There will be a
small inaccuracy when $\omega r_s/c$ exceeds 0.45 and significant
inaccuracy is expected when $\omega r_s/c > 1$.

Using the formulations in \cite{Ye3,AAD2} for acoustic
transmission and using the formulation in \cite{Kushuwaha} for
computing the acoustic band structure, we examine the results in
Fig.~3 of \cite{Kaf}. It should be noted that the formulation in
\cite{Ye3} only takes into account the pulsating mode, while the
formulation in \cite{AAD2} incorporates all possible vibrational
modes of the bubbles. The two approaches agree well except at
higher resonance frequencies of $\omega r_s/c \approx 0.48$ and
$0.77$\cite{Kaf}. Thus, in the following, the approach in
\cite{Ye3} will be used.

Employing the same parameters as in \cite{Kaf}, we reproduce
partial results in Fig.~1. For comparison, we show the result for
a single random configuration (a) and as well as the ensemble
averaged result (b). It is clear that whether making ensemble
average makes not much difference for $\omega r_s/c_0 < 0.6$. In
particular, the ensemble average basically plays no role for
frequencies located around the complete band gap. Compared to
Fig.~3 in \cite{Kaf}, there are several differences. (1) The
complete band gap is a little wider in the present case, but is
consistent with a transmission calculation, following \cite{Ye4}.
To further inspect the difference, we have also computed the band
structures at the air void fraction of $10\%$, yielding nearly the
same answer as in Fig.~1(a) of \cite{Kaf}. (2) In contrast to
\cite{Kaf}, the transmission in the present computation changes
sharply at the frequencies slightly above the natural frequency of
an air-bubble in water\cite{Morse}, which in fact leads to a first
order phase transition from the extended state to the localized
state\cite{Ye4}. Exploring the difference, we have considered
possible effects of the locations of transmitting source and
receiver on the transmission, and find that the transmission is
insensitive to the positions of the transmitter and the receiver
for frequencies between $\omega r_s/c_0 = 0.0136$ and about $0.3$,
indicating that the discrepancy in the transmission behavior
between the present results and that from \cite{Kaf} is not caused
by a difference in the source or receiver location. The
transmission at frequencies higher than roughly $\omega r_s/c_0 =
0.3$, however, is indeed sensitive to the location of either
transmitter or receiver. (3) In Fig.~3 of \cite{Kaf}, the range of
inhibited transmission coincides with the band gap, while in our
case this is not the case. In addition, our results show that the
randomness tends to smear out the transmission peak near the upper
band edge, consistent with 2D cases\cite{Ye8}. At first sight, the
present inconsistency between the range of the band gap and the
range of the inhibited transmission seems to suggest errors in our
computation. But, in fact, the the disagreement is due to the
effect of the finite sample size. To elaborate, Fig.~2 shows the
results for different sizes of bubble clouds, where the effect of
finite sample size is obvious. The range of the transmission
inhibition shall agree with the band gap when the sample size
tends infinity.

When the multiple scattering is switched off, the effect of
ensemble average becomes more prominent, comparing the dotted
lines in the right panels in Fig.~1(a) and (b). Again we see
discrepancies with Fig.~3 in \cite{Kaf}. For example, the present
result for the lattice case starts to flat out at about $\omega
r_s/c_0 = 0.3$, while that in \cite{Kaf} at about 0.4. The
randomness tends to broad the transmission peak located around the
natural frequency of the breathing mode of the air-bubble. The two
transmission peaks at $\omega r_s/c_0 = 0.48$ and $0.77$ shown in
\cite{Kaf} are absent from the right panel of Fig.~1. This is
because the results in Fig.~1 did not take into consideration
higher vibrational modes. When the higher modes are considered
using \cite{AAD2}, the overall results in the right panel of
Fig.~1 will not change much, except that two sharp transmission
peaks will appear as in Fig.~3(c) of \cite{Kaf}.

Discussion with Dr. Pigang Luan is greatly appreciated. This work
received support from National Science Council (Grant No.
NSC-89-2112-M008-008 and NSC-89-2611-M008-002).


\begin{description}

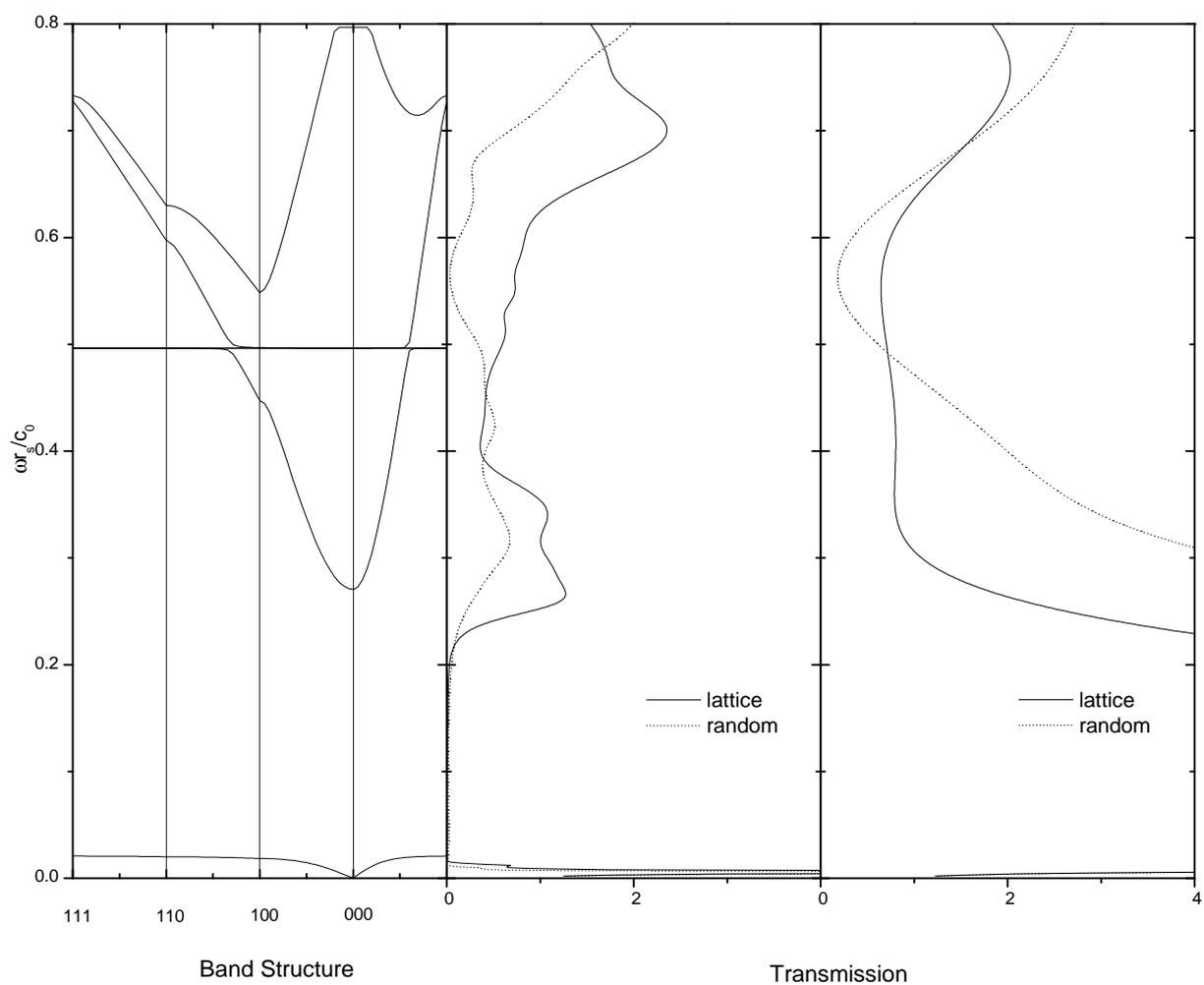
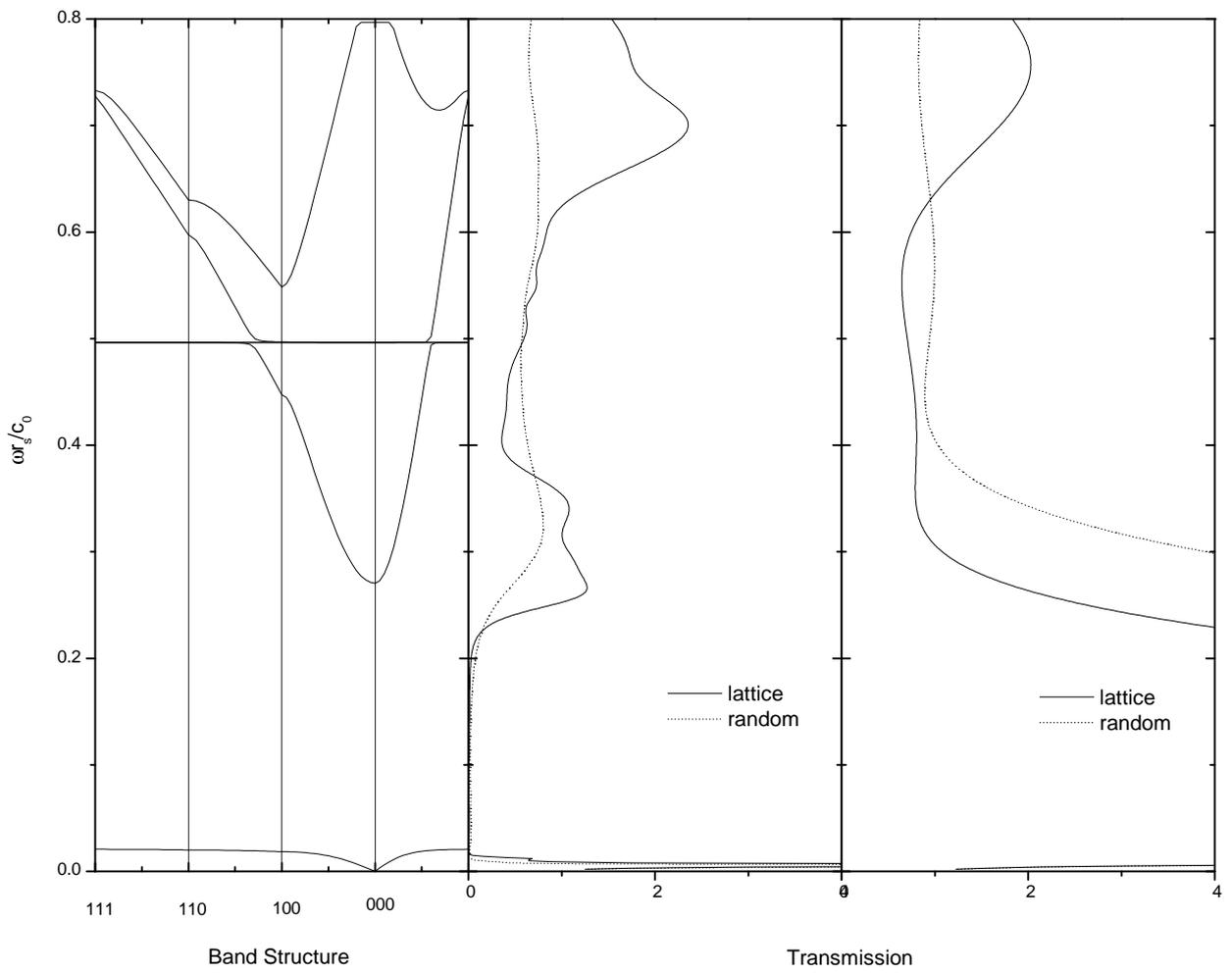
\item[Figure 1] (a) {\bf Left panel}. Acoustic bands in a simple cubic array of
air-bubbles in water. $r_s$ is the bubble radius, $\omega$ is the
angular frequency and $c_0$ is the water velocity in the water.
Air volume fraction is $1\%$. {\bf Middle panel}. Reduced
transmission coefficient through the bubble array in ordered
(solid line) or one random (dotted line) placement.  {\bf Right
panel}. The same as in the middle panel, but with the multiple
scattering switched off. (b) The same as in (a), but the
transmission in the case of random placement is averaged over 100
random configurations.

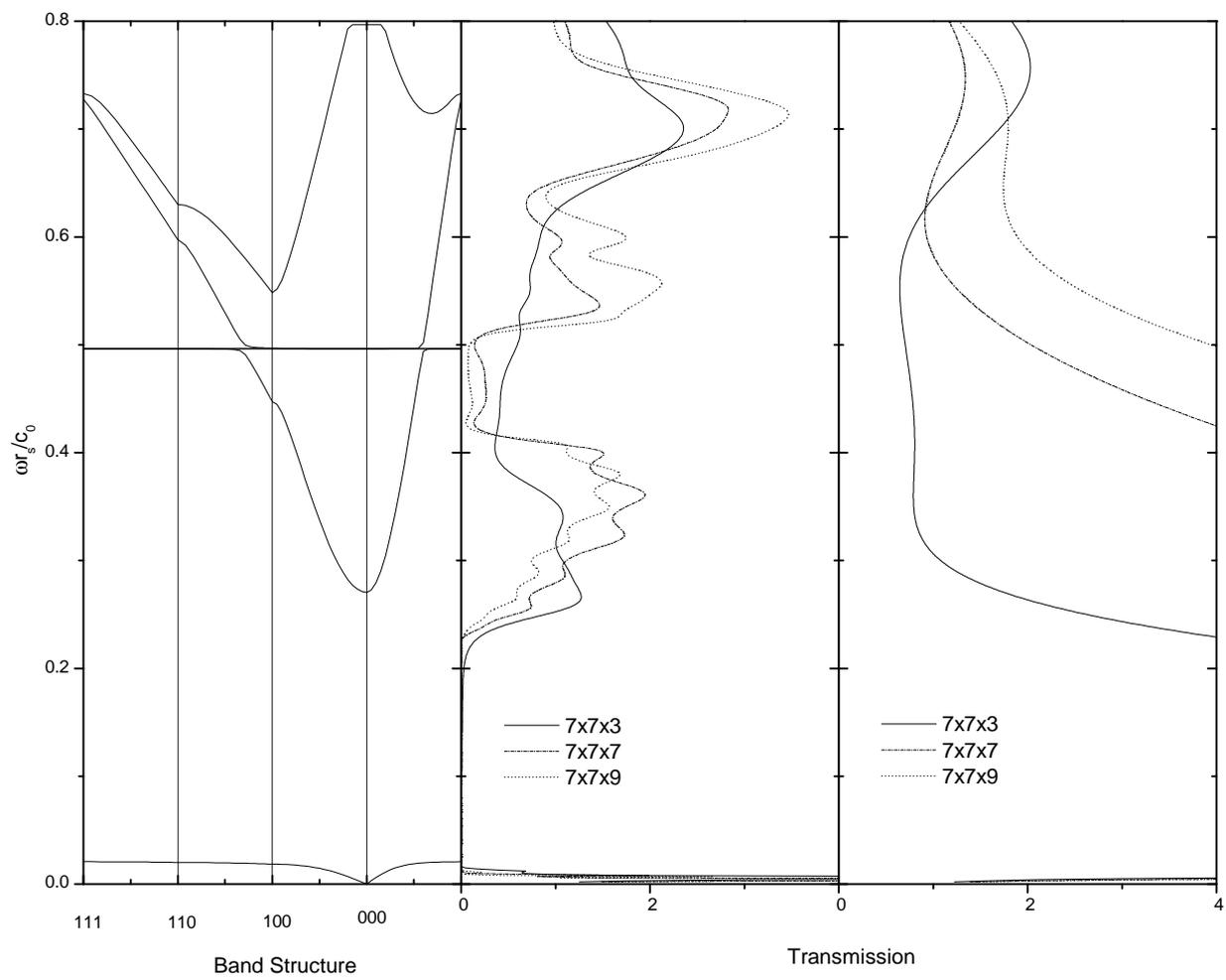
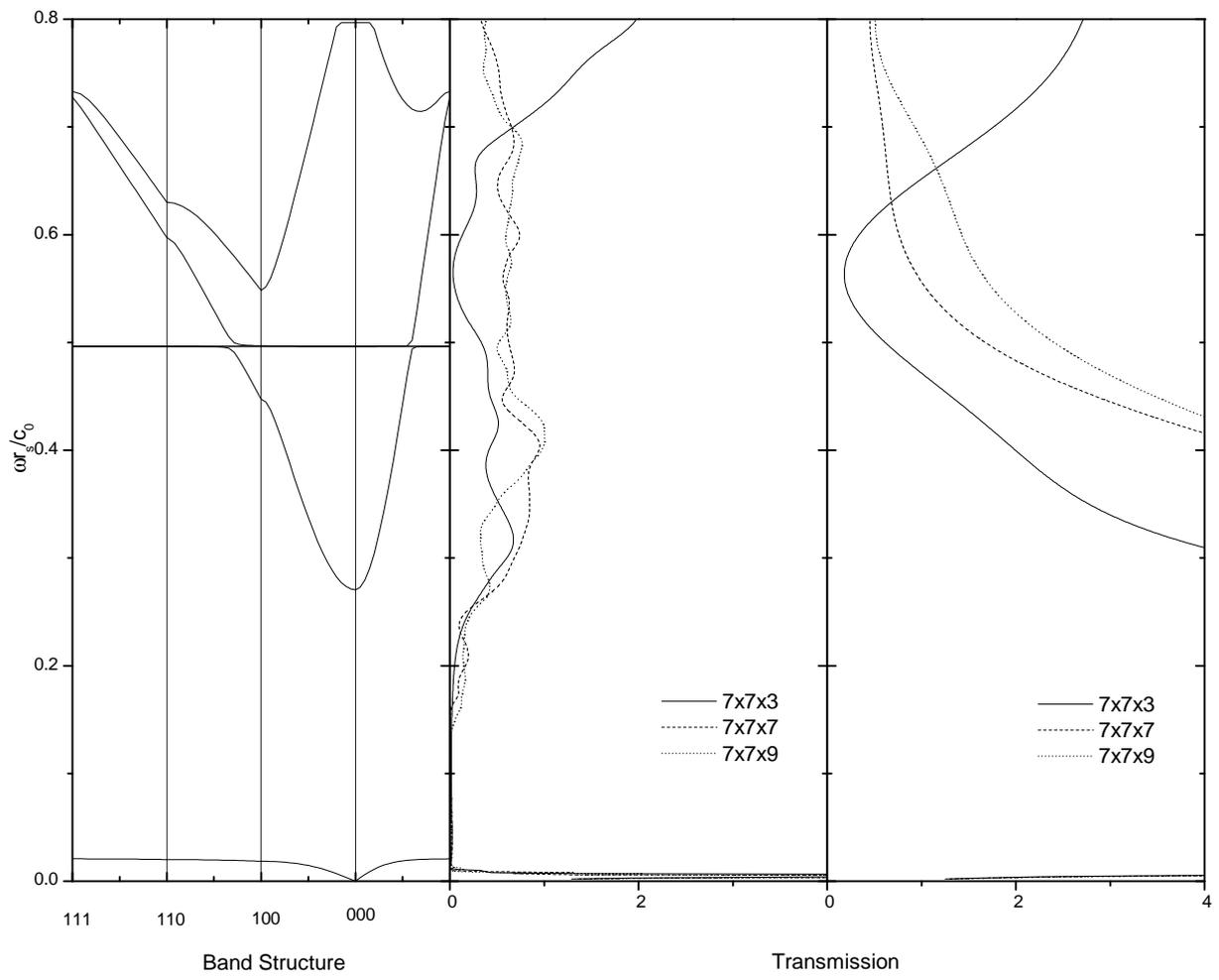
\item[Figure 2] The same as in Fig.~1, but for various sizes of
bubble clouds in ordered placement (a) and as well as for one
random arrangement (b).

\end{description}

\end{document}